\documentclass{revtex4}

\usepackage{amsmath}
\usepackage[pdftex]{graphicx}     

\begin{document}

\title{Cylindrical Magnets and Ideal Solenoids}%
\author{Norman Derby$^{a)}$}
\affiliation{Community College of Vermont, Bennington, VT 05201}
\author{Stanislaw Olbert$^{b)}$}
\affiliation{Department of Physics (Emeritus), Massachusetts Institute of Technology, Cambridge, MA 02139}

\begin{abstract}
Both wire-wound solenoids and cylindrical magnets can be approximately modeled as ideal, azimuthally symmetric solenoids.  We present here an exact solution for the magnetic field of an ideal solenoid in an especially easy to use form. The field is expressed in terms of a single function that can be rapidly computed by means of a compact, highly efficient algorithm, which can be coded as an add-in function to a spreadsheet, making field calculations accessible even to introductory students. In computational work these expressions are not only accurate but also just as fast as most approximate expressions. We demonstrate their utility by numerically simulating the experiment of dropping a cylindrical magnet through a nonmagnetic conducting tube and then comparing the calculation with data obtained from experiments suitable for an undergraduate laboratory.
\end{abstract}

\maketitle
\section{Introduction.}
\label{sec: intro}
The following article has been accepted by the American Journal of Physics. After it is published, it will be found at http://scitation.aip.org/ajp/.

Solenoids and cylindrical magnets are staples of introductory physics laboratory experiments and demonstrations. When it comes time to put theory to the test, simple models for these objects are needed. An idealized solenoid –-- a solenoid with strictly azimuthal current in a thin sheet wrapped around a right circular cylinder –-- can serve as a reasonable model of an actual wire-wound solenoid, and can serve as an even better model of a permanent cylindrical magnet, provided that its magnetization is sufficiently uniform.

At the introductory level, the magnetic field of an ideal solenoid of finite length can only be computed exactly along the symmetry axis, where the field can be expressed in terms of elementary functions. At off-axis points, geometry makes things more difficult and introductory-level students generally have no tools for obtaining even approximate values for the field except at very large distances where the field resembles that of a point dipole. This surely makes it harder for students to develop familiarity and confidence in dealing with magnetic phenomena.

It is well known that the field due to a circular current loop can be written in terms of elliptic integrals, so by treating the ideal solenoid as a stack of loops, its magnetic field can be obtained by a straightforward integration. Alternatively, the field may be derived by solving a boundary value problem with cylindrical symmetry. In either case, exact expressions for the field have been known for over a century. They can be expressed in different forms using various special functions such as elliptic integrals, Heuman's lambda function, various Bessel functions, hypergeometric functions, etc.\cite{Conway2001, Labinac, Garrett1, Garrett2, Callaghan}

It is true that some of these expressions can be rather cumbersome looking, and there are certainly many situations in which an approximate expression may provide better physical insight, be easier and/or less error prone to use, or serve some higher pedagogical purpose. However, when the expression for the magnetic field is merely being used as part of a computation of another quantity, it would seem that the only serious question about using an exact solution would be its computational speed.

However, we easily found nearly a dozen recent papers\cite{Pelesko, Roy, Horton, Hahn, Pissaro, Knyazev, Iniguez, Amrani, Levin, MacL, question} (see section \ref{subsec:previous}) that included computations involving the field of a cylindrical magnet but did not make use of exact solutions in any form.  The reasons for this choice probably varied from one paper to the next, but several papers explicitly stated that the theoretical treatment of the magnetic field due to a solenoid of finite length was too complicated for their purposes.

There are apt to occur many occasions in which calculations involving solenoid fields arise. As a convenient tool for such situations, we present an exact solution in a form that  (1) is algebraically less complicated, (2) does not require any previous knowledge of special functions, and (3) comes with a numerical algorithm that is simple and efficient. The field is expressed in terms of a single function, a generalized complete elliptic integral. This function is completely defined by an integral whose form occurs naturally in problems involving cylindrical symmetry. Numerical values can be computed by means of an algorithm that can be easily coded on even a programmable calculator or employed as a user-defined function or macro in a spreadsheet. In this form, the finite-length, ideal solenoid model is as simple and fast to use as the point dipole model in computations.

This paper will present these exact expressions for the magnetic field of an ideal solenoid and exact expressions for its self-inductance, and will then provide a brief illustration of the effectiveness of these expressions by simulating the experiment of dropping a cylindrical magnet through a nonmagnetic conducting tube and comparing the calculations with the results of some simple experiments.
\section{The Generalized Complete Elliptic Integral.}
\label{sec: cel}
In computing fields with cylindrical symmetry, certain integrals occur in a natural way. They are special cases of a function defined by a generalized complete elliptic integral ($cel$):
\begin{equation}
\label{celdef}
C(k_c ,p,c,s) = \int\limits_o^{\pi /2} {\frac{{c\cos ^2 \varphi  + s\sin ^2 \varphi }}{{(\cos ^2 \varphi  + p\sin ^2 \varphi )\sqrt {\cos ^2 \varphi  + k_c^2 \sin ^2 \varphi } }}\,} d\varphi .
\end{equation}
\rm Appendix~\ref{sec: appcel} describes code for an extremely efficient numerical algorithm for calculating values for $C$. The example code is presented in a dialect of BASIC that can be directly used as a user-defined function in spreadsheet program or can easily be translated to other programming languages. This appendix also contains further information about $C$, including the relationship of $C$ to other forms of elliptic integrals and links to code.
\section{Magnetic Field Expressions}
\label{sec: field}
Consider a cylinder of length $2b$ and radius $a$ wrapped by an azimuthal sheet of current $I_{total}$, equivalent to a tightly wound solenoid with a number of turns per unit length $n$ carrying a current $I$,  i.e. $I_{total} = 2bnI$. The magnetic moment $\mu$ of the solenoid is $\mu = 2bnI \cdot \pi a^2 $. It is well known that along the symmetry axis of a such a solenoid the field, using cylindrical coordinates ($\rho ,\varphi, z$) with the origin at the center of the solenoid, takes the form
\begin{equation}
\label{finsolax}
B_z = \frac{{\mu _o nI}}{2}\left\{ {\frac{{z + b}}{{\sqrt {\left( {z + b} \right)^2  + a^2 } }} - \frac{{z - b}}{{\sqrt {\left( {z - b} \right)^2  + a^2 } }}} \right\} ,
\end{equation}
which reduces to $B_z = \mu _o nI$ for an infinite solenoid. For the general case, (see Appendix~\ref{sec: appder} for an outline of the derivation) the magnetic field components are:
\begin{equation}
\label{Brho}
B_\rho   = B_o \left[\, {\alpha _ +  \,C\left( {k_ +  ,1,1, - 1} \right) - \alpha _ -  \,C\left( {k_ -  ,1,1, - 1} \right)} \, \right]
\end{equation}
and
\begin{equation}
\label{Bz}
B_z  = \frac{{B_o \, a}}{{a + \rho }}\left[ \,  { \beta _ +  \,C\left( {k_ +  ,\gamma^2 ,1,\gamma} \right) - \beta _ -  \,C\left( {k_ -  ,\gamma^2 ,1,\gamma} \right)} \, \right] ,
\end{equation}
with
\begin{equation}
\label{Bo}
B_o  = \frac{{\mu _o }}{\pi }nI,
\end{equation}
\begin{equation}
\label{zpm}
z_ \pm   = z \pm b ,
\end{equation}
\begin{equation}
\label{alphapm}
\alpha _ \pm   = \frac{a}{{\sqrt {z_ \pm ^2  + (\rho  + a)^2 } }} ,
\end{equation}
\begin{equation}
\label{betapm}
\beta _ \pm   = \frac{z_ \pm}{{\sqrt {z_ \pm ^2  + (\rho  + a)^2 } }} ,
\end{equation}
\begin{equation}
\label{s}
\gamma = \frac{{a - \rho }}{{a + \rho }} ,
\end{equation}
\begin{equation}
\label{kpm}
k_ \pm  = \sqrt {\frac{{z_ \pm  ^2  + \left( {a - \rho } \right)^2 }}{{z_ \pm  ^2  + \left( {a + \rho } \right)^2 }}} .
\end{equation}
These compact forms involve only a single function, $C$. They compute quickly and accurately both inside and outside the solenoid and are mathematically well-behaved except on the edge of the current sheet at $\rho = a$ and $z = \pm b$.

\begin{figure}
  \begin{center}
    \includegraphics[scale=.6]{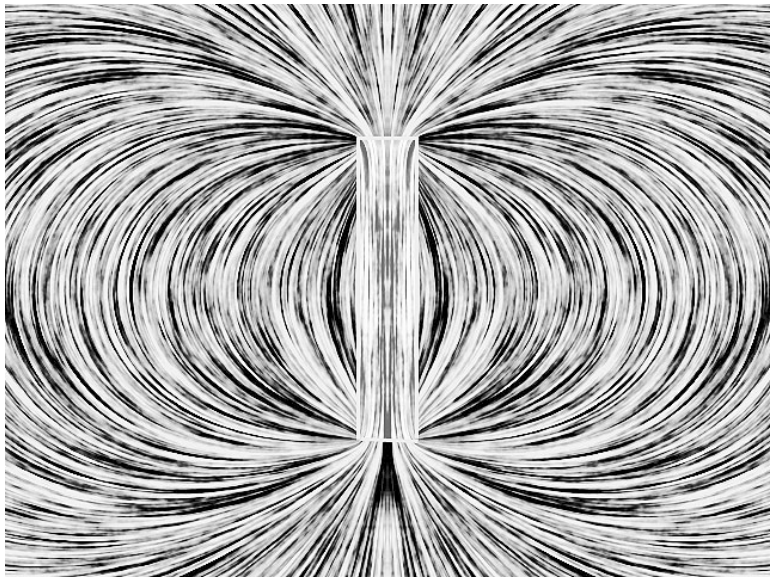}
  \end{center}
  \caption{Field lines of an ideal solenoid with a length that is five times its diameter. }
  \label{fig:DLIC}
\end{figure}

These expressions reveal that, if distances are measured in units of $a$, then the magnetic field lines of an ideal solenoid depend only upon the ratio of length to diameter, $b/a$. Fig.~\ref{fig:DLIC} shows the structure of field lines for a solenoid with $b = 5 a$ using the line integral convolution method employed by Sundquist\cite{Sundquist} and Belcher.\cite{Belcher}  Using the above expressions for the field, this image was produced by their JAVA program in a matter of seconds. The behavior of field lines within the solenoid contrasts sharply with that of external field lines and indicates why a single approximate formula in terms of elementary functions has difficulty representing the field at both near and far distances.
\subsection{Limits}
Along the axis of the solenoid ($\rho = 0$), $k_ \pm = \gamma = 1$ and $C(1,1,1,1) = \pi/2$, so Eq.~\eqref{Bz} reduces to Eq.~\eqref{finsolax}.

As $b \to 0$ with $2bnI = I_{total}$ remaining finite, a solenoid becomes a current loop and the field expressions above with $0 < b << a$ do approximate those of a current loop. In the limit $b=0$ they take the form:
\begin{equation}
\label{Brholoop}
B_\rho   = \frac{{\mu _o }}{\pi }\frac{{I_{total} az}}{{\left[ {z^2  + \left( {\rho  + a} \right)^2 } \right]^{{\raise0.7ex\hbox{$3$} \!\mathord{\left/
 {\vphantom {3 2}}\right.\kern-\nulldelimiterspace}
\!\lower0.7ex\hbox{$2$}}} }}C\left( {k_1 ,k_1 ^2 , - 1,1} \right)
\end{equation}
and
\begin{equation}
\label{Bzloop}
B_z  = \frac{{\mu _o }}{\pi }\frac{{I_{total} a\left( {\rho  + a} \right)}}{{\left[ {z^2  + \left( {\rho  + a} \right)^2 } \right]^{{\raise0.7ex\hbox{$3$} \!\mathord{\left/
 {\vphantom {3 2}}\right.\kern-\nulldelimiterspace}
\!\lower0.7ex\hbox{$2$}}} }}C\left( {k_1 ,k_1 ^2 ,1,\gamma } \right) ,
\end{equation}
where
\begin{equation}
\label{k1}
k_1 ^2  \equiv \frac{{z^2  + \left( {a - \rho } \right)^2 }}{{z^2  + \left( {a + \rho } \right)^2 }}.
\end{equation}

Finally, at large distances from the solenoid ($r >> a,b$), the field reduces to that of a point dipole:
\begin{equation}
\label{dipole}
B_\rho   = \frac{{\mu _o \mu }}{{4\pi }}\frac{{3\rho z}}{{r^5 }},\,\,\,\,B_z  = \frac{{\mu _o \mu }}{{4\pi }}\frac{{\left( {2z^2  - \rho ^2 } \right)}}{{r^5 }}
\end{equation}
with
\begin{equation}
\label{r}
r^2  = \rho ^2  + z^2 .
\end{equation}

\subsection{Inductance}
It is worth noting that it is also possible to derive an exact expression for the mutual inductance of two coaxial ideal solenoids.\cite{Grover, Olshausen, Garrett3} The self inductance $L$ of a solenoid can then be obtained as a special case. In our notation the self inductance can be expressed very compactly:
\begin{equation}
\label{L}
L = \frac{8}{3}\mu _o \left( {na} \right)^2 \left[ {\sqrt {a^2  + b^2 } \, C\left( {k_o ,1,1,2k_o ^2 } \right) - a} \right] ,
\end{equation}
where
\begin{equation}
\label{ko}
k_o  = \frac{b}{{\sqrt {a^2  + b^2 } }}.
\end{equation}
\section{Falling Magnets}
\subsection{Previous Work}
\label{subsec:previous}
Faraday's law is often dramatically demonstrated by dropping a small,
highly magnetized, cylindrical permanent magnet (radius $a$, length $2b$, mass $m$, magnetic moment $\mu $) into a
vertical, non-magnetic tube of conductivity $\sigma$, relative permeability 1,
length $L$, inside radius $r$ and the wall thickness be $w << r$).\cite{Hahn, Amrani, Levin, Clack, reed, video}
The small masses and large magnetic moments of rare earth magnets give them a startling ``hang'' time. After the initial surprise subsides, students begin to ask questions. ``How does the time of fall depend on the diameter or conductivity of the tube?'' ``How does it depend on the length of the magnet?''

This experiment or similar ones have been analyzed in several papers.
Pelesko {\it et al.} \cite{Pelesko} and Roy {\it et al.} \cite{Roy} use dimensional analysis to show that
in a thin-walled tube the speed of the falling magnet should be proportional to $(mgr^4)/(\sigma \mu^2 w)$, which demonstrates the power of dimensional analysis, though it is unable to provide information about how the speed might depend upon the geometry of the magnet.
Other papers employ Faraday's law, but differ in the ways they model the cylindrical magnet.
Hahn {\it et al.} \cite{Hahn} derive Eq.~\eqref{Fdrag} for the force on a magnet oscillating in a tube but then evaluate this by treating the magnet as a point dipole.
Knyazev {\it et al.} \cite{Knyazev} treat only the point dipole case, but expand the analysis to include high speeds not attainable in demonstrations.
Levin {\it et al.} \cite{Levin} initially note that the ideal solenoid's external field is exactly equivalent to that of two uniformly (magnetically) charged discs at the top and bottom of the magnet, but then use point monopoles instead of discs in their calculations.
I\~niguez {\it et al.}  \cite{Iniguez} model the interaction of the magnet and the tube by means of an elaborate equivalent resistor network and provide a sample calculation in which the magnet is treated as a point dipole.
Calculations based on the dipole approximation do not predict experimental results with much accuracy when the magnet fits closely within the tube, especially when the length of the magnet is greater than its diameter.

After deducing Eq.~\eqref{vterm} below, MacLatchy {\it et al.}  \cite{MacL} model the cylindrical magnet as a stack of several polygonal loops and then compute its field from the Biot-Savart law. Although slightly cumbersome, this approach does offer adjustable accuracy.
Partovi and Morris \cite{Partovi} offer a comprehensive treatment of a cylindrical magnet moving at an arbitrary, nonrelativistic velocity in an infinite tube of arbitrary thickness and permeability. This is a boundary-value problem with cylindrical symmetry that they solve, expressing the drag force on the magnet in terms of integrals involving Bessel functions with complex arguments. Their results are exact, though restricted to a steady state situation. The integral expressions in their paper are indeed daunting to look at, but the authors provide sample Mathematica code for computing them.

The expressions for magnetic field presented in section \ref{sec: field} are extremely well suited to this problem.
\subsection{Theory}
Previously cited references \cite{Hahn, MacL, Partovi} discuss the theory in some detail, so we only provide a quick sketch here.  Choose cylindrical coordinates with the $z$-axis vertical and the origin located (momentarily) at the center of the magnet. In an experimental situation, the conductivity of the tube can be determined by measuring its resistance per unit length $R_L$:
\begin{equation}
\label{RL}
R_L = \frac{1}{{\sigma 2 \pi w \bar r }} ,
\end{equation}
where $\bar r = r + w/2 $.

\begin{figure}
    \begin{center}
    \includegraphics[scale=0.6]{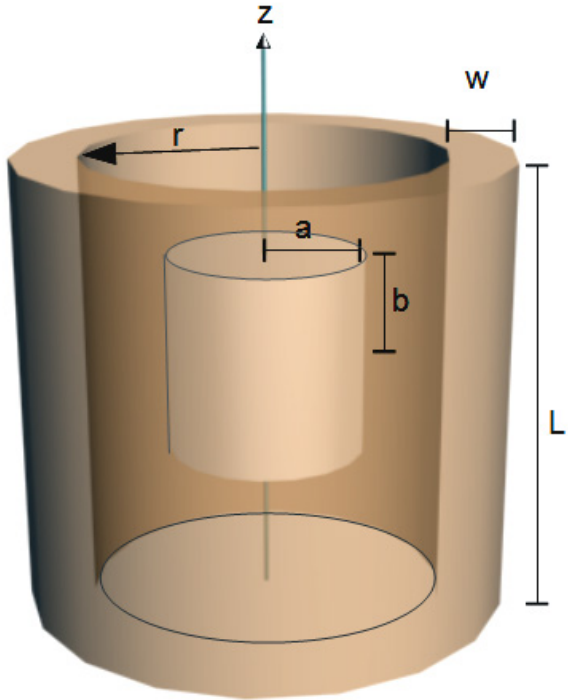}
    \end{center}
    \caption{Geometry for a magnet magnet falling though a non-magnetic, conducting tube. }
    \label{fig:tubegeo}
\end{figure}

As the magnet falls, the changing magnetic field within the tube walls is accompanied by an electric field
that drives currents that cause ohmic heating. Although currents in the tube in turn induce currents
within the permanent magnet, it is easy to show that under our experimental conditions the
only significant energy losses are those within the tube walls. The speed of fall is so slow
that air resistance is also quite negligible. In the following discussion, we assume that the magnet fits closely enough within the tube walls that its axis remains vertical and cylindrical symmetry is maintained during the fall. (When a small diameter magnet falls within a much larger diameter tube, one observes the axis of the magnet precessing about the vertical during the fall.)

The electric field within the tube can be deduced by arguing that, in the reference frame of the falling magnet, there is only a magnetostatic field, so in the frame of reference of the tube, where the magnet has velocity ${\bf v}$, there must be an electric field
${\bf{E}} = - {\bf{v} \times \bf{B}}$,  or  $E_\varphi = - v_z B_\rho$.
Alternatively, the field can be obtained from Faraday's law by considering a horizontal circular loop of radius $\bar r$ lying within the tube walls at height $z'$ above the center of the
magnet. Defining the upward magnetic flux though such a loop as
\begin{equation}
\label{upflux}
\Phi \left( z' \right) = \int\limits_0^{\bar r} {B_z \left( {\rho ,z'} \right)} 2\pi \rho d\rho ,
\end{equation}
the $emf$ around the loop is
\begin{equation}
\label{emf}
{emf} =  - \frac{{d\Phi }}{{dt}} =  - \frac{{d\Phi \left( z' \right)}}{{dz'}}\frac{{dz'}}{{dt}} = v_z\frac{{d\Phi \left( z \right)}}{{dz}}.
\end{equation}
since
\begin{equation}
\label{vanddzdt}
v_z =  - \frac{{dz' }}{{dt}}.
\end{equation}

If you then take two such loops separated by a vertical distance $dz$, you can visualize them as the edges of a small cylindrical Gaussian pill box. Since there are no magnetic monopoles, the total magnetic
flux leaving the closed surface of the box must be zero:
\begin{equation}
\label{nomonopoles}
\Phi \left( {z + dz} \right) - \Phi \left( {z} \right) + B_\rho  \left( {\bar r,z} \right)2\pi \bar rdz = 0 = d\Phi  + B_\rho  \left( {\bar r,z} \right)2\pi \bar rdz .
\end{equation}
Combining this with Eq.~\eqref{emf}, we have as before
\begin{equation}
\label{ephi}
{emf} =  - \left( {2\pi \bar r\,v_z} \right)B_\rho   = E_\varphi  2\pi \bar r .
\end{equation}

The force acting on the falling magnet can be deduced from energy considerations. The electric field within the tube drives currents that dissipate energy at a rate per unit volume of $\sigma E_\varphi  ^2$,
so the total power lost $P$ can be obtained by integrating this over the volume of the walls.
If the walls are thin, the power loss when the magnet is at height $z$ above the bottom of the
tube is
\begin{equation}
\label{heat}
P = \int {\sigma E_\varphi  ^2 2\pi w\bar r\,dz}  = 2\pi w\bar r\sigma v_z ^2 \int_{ - z}^{L - z} {B_\rho  ^2 \left( {\bar r,z'} \right)dz'} = - v_z F_{drag}.
\end{equation}

Alternatively, the Lorentz forces acting on the currents can be calculated directly. A horizontal slice of tubing of height $dz$ located at height $z$ above the center of the magnet is a circuit with electrical resistance
\begin{equation}
\label{RE}
R_E  = \frac{1}{\sigma }\frac{{2\pi \bar r}}{{w\,\,dz}},
\end{equation}
so the current due to the $emf$ around the ring is
\begin{equation}
\label{current}
dI = \frac{emf}{R_E} =  - \sigma v_z w\,B_\rho  \,dz .
\end{equation}
The vertical force exerted on this current by ${\bf B}$ is:
\begin{equation}
\label{Fz}
dF_z  = -2\pi \bar r\,dI\,\,B_\rho   =   \sigma \,2\pi \bar rw\,v_z B_\rho ^2 \,dz ,
\end{equation}
and the force on the magnet then follows from Newton's third law. The total vertical force on the magnet is the sum:
\begin{equation}
\label{Fdrag}
F_{drag}  =  -  \frac{v_z}{R_L} \,\,\int_{ - z}^{L - z} {B_\rho  ^2 \left( {\bar r,z'} \right)dz'} .
\end{equation}
in agreement with Eq.~\eqref{heat}.

The magnetic field computes so quickly using Eq.~\eqref{Brho} for $B_{\rho}$ that this drag force can be used to numerically integrate the equation of motion for the falling magnet even though an integration must be performed at each time step.

The powerful but light-weight magnets used in demonstrations reach a constant velocity within a fraction of a second. Since the magnitude of $B_\rho$ declines extremely rapidly with $z$, this terminal velocity can be calculated by equating $F_{drag}$ to $mg$ and, with little error, setting the limits of integration to $\pm \infty$:
\begin{equation}
\label{vterm}
v_{terminal}  = \frac{{ - mg R_L}}{{\int_{ - \infty }^\infty  {B_\rho  ^2 \left( {\bar r,z'} \right)dz'} }} .
\end{equation}
\subsection{Experiment}
\label{sec: experiment}
We checked these predictions experimentally in the simplest possible way using equipment and procedures available in a typical undergraduate laboratory.
\begin{table} [ht]
\begin{center}
\begin{tabular}{|c|c|}
\hline
\bf{Physical}  &  \bf{Electrical} \\
\hline
L = 1.478 m  &   length between clips = 1.475 m\\
\hline
r = 7.25 mm  &  I = 4.95 A\\
\hline
w = 0.7 mm  &   $\bigtriangleup V = 4.49 mV$  \\
\hline
mass = 434 g  &   $\sigma$ = $56.0 \times 10^6$ S/m \\
\hline
         & $R_L$ = $5.37 \times 10^{-4}\,\Omega/m$\\
\hline
\end{tabular}
\caption{Physical and electrical properties of tube. Conductivity was determined by attaching a power supply to the ends of the tube with alligator clips, running a current $I$ through the tube, and measuring the potential drop $\bigtriangleup V$. }
\label{table:tubeandmag}
\end{center}
\end{table}
We used copper plumbing tube (Table \ref{table:tubeandmag}), which is about 99\% pure copper.  Its magnetic permeability was not measured but taken to be the same as pure copper. The resistance per unit length was determined by running a current of several amperes through the tube while measuring a few millivolts potential difference across it. The conductivity of pure copper is about $110\%$ of the standard IACS value, $5.8108 \times 10^7$ S/m at $20\,^{\circ}\mathrm{C}$, while the conductivity of typical copper tubing for plumbing is typically about $85\%$ of the IACS value.

We obtained six cylindrical magnets with various lengths but with the same half-inch ($12.7$ mm) diameter. We determined the magnetic moment $\mu$ for each one by using a small magnetometer to measure the field strength at several points along its axis and then adjusting the value of ($nI$) in Eq.~\eqref{finsolax} to give a best fit.

Finally, holding a magnet vertically by its upper 4 millimeters, we inserted it into the top of the tube and released it while manually starting a timer that stopped when the bottom  millimeter of magnet activated a photogate placed at the bottom of the tube. The results are summarized in Table~\ref{table:dropdata}.
\begin{table} [ht]
\begin{center}
\begin{tabular}{|c|c|c|c|c|c|c|c|}
\hline
No. & b/a & m(g) & $\mu (A \cdot m^2)$ & $v_{average}$ (m/s) & $v_{t}$-Partovi & $v_t$-Eq.~\eqref{vterm} & $V_{average}$  \\
\hline
1  &     1.0  &    12.1  &    1.76  &     0.0687  &     0.0669  &     0.0670  & 0.0674\\
\hline
2  &     1.5  &    17.9  &    2.36  &     0.1045  &     0.1050  &     0.1050  & 0.1058\\
\hline
3  &     2.0  &    23.8  &    3.23  &     0.1275  &     0.1243  &     0.1243  & 0.1254\\
\hline
4  &     3.0  &    36.4  &    5.00  &     0.1825  &     0.1710  &     0.1711  & 0.1731\\
\hline
5  &     4.0  &    48.2  &    6.37  &     0.2473  &     0.2451  &     0.2451  & 0.2486\\
\hline
6  &     1.0  &    12.9  &    1.17  &     0.1513  &     0.1615  &     0.1616  & 0.1622\\
\hline
\end{tabular}
\caption{Six cylindrical magnets with half-inch diameters and various ratios of length to diameter ($b/a$) dropped from rest through a length of copper tube. Average speed of fall is compared to predictions of terminal velocity using formula of Partovi and Morris or Eq.~\eqref{vterm}. Last column is the predicted average velocity obtained by numerically integrating the equations of motion of a magnet experiencing the drag force given by Eq.~\eqref{Fdrag}. }
\label{table:dropdata}
\end{center}
\end{table}
In addition to the measured average velocity of each falling magnet, the table shows pairs of terminal velocities numerically calculated both from Eq.~\eqref{vterm} and from the lengthy exact expression for the terminal velocity in Partovi and Morris. (Both calculations were carried out in Python using the SciPy libraries.) The numerical agreement between these two different ways of calculating terminal velocity validates the approximations used in deriving Eq.~\eqref{vterm}.

\begin{figure}
    \begin{center}
     \includegraphics[scale=0.5]{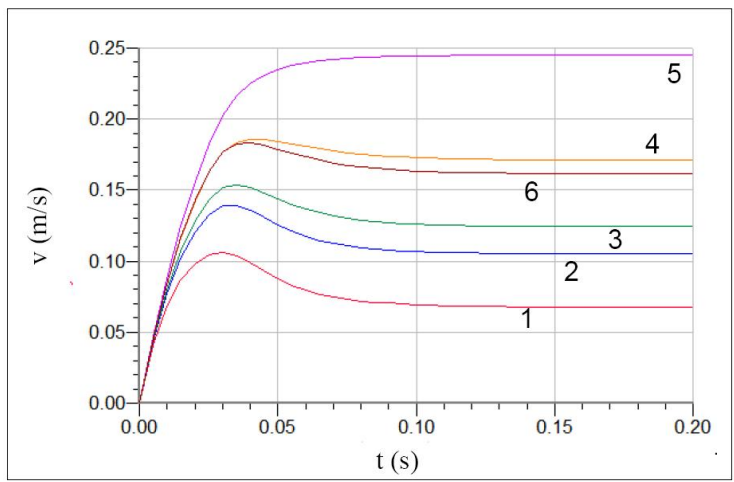}
    \end{center}
    \caption{Speeds of magnets dropped through a copper tube. Each magnet was inserted into the tube and then released from rest. These are numerical solutions of the equation of motion including the drag force during the first fifth of a second after release.   }
    \label{fig:initialfall}
\end{figure}

In addition to calculating terminal velocities, we used Eq.~\eqref{Fdrag} for the drag force on a falling magnet to numerically solve the equations of motion for $z(t)$ and thus the total time of fall. Because of their differing lengths, the center of each magnet started out at a different height in the tube resulting in slight differences in their initial behaviors, which are plotted in Fig.~\ref{fig:initialfall}, which shows that in all cases terminal velocity was achieved within a few tenths of a second resulting in terminal speeds that are only slightly smaller than average speeds of fall.
Since the experimental procedures were not very sophisticated, the excellent agreement between theory and experiment was surprising. The experiment is clearly within the capabilities of introductory students. How much of the numerical simulation would be appropriate to ask them to do? That depends upon the level of the class and the amount of time allotted to this project. It is possible, though, to perform the computation of the terminal velocity at an introductory level if students are comfortable with spreadsheet programs or elementary programming concepts in, for example, VPython. Three things permit this. First, $cel$ can be added to a spreadsheet as a user-defined function (see Appendix~\ref{sec: appcel}), so $B_{\rho}$ can be computed and graphed. Second, $B_{\rho}$ is a symmetrical function of $z$, so the integral in $v_{terminal}$ only needs to be evaluated from $0$ to $\infty$. And third, at large enough values of $z$, $B_{\rho}$ approaches the dipole form (Eq.~\eqref{dipole}), so $B_{\rho}^2 \sim z^{-8}$, allowing the integral to be truncated at relatively small values of $z$. In fact, students can plot $B_{\rho}^2$ in a spreadsheet and see that it starts at $0$ at $z=0$, reaches a sharp maximum near $z=b$, and decreases by several orders of magnitude by $z=3b$. As a result, the entire integral may be computed as a simple Riemann sum from $z=0$ to about $z=3b$ using roughly a step size $dz \simeq 0.01 b$ -- for which a spreadsheet is an ideal tool. Any concern about truncation errors could be addressed by calculating the remainder of the integral analytically using the dipole approximation for $B_{\rho}$. More advanced students, of course, might be acquainted with other programs or numerical methods for handling these tasks, but the elementary method described gives results within about a percent of the more sophisticated tools and can easily be improved by using more steps and/or smaller size steps.
\section{Comments}
We have presented very compact expressions for the magnetic field of an ideal solenoid and have demonstrated that in numerical work they are easy to use, remarkably fast and can readily be incorporated into spreadsheet calculations. This makes it possible to easily simulate a variety of situations involving cylindrical magnets. In the case of the falling magnet demonstration, we have shown that a simple treatment of the problem provides results that agree with those from a more complicated analysis and are consistent with simple measurements. It should be clear that the methods used here can be applied to other cylindrically symmetric situations such as the electric fields of uniformly charged rings or cylindrical shells. Such fields can then be written in terms of $cel$ conferring upon calculations all the advantages that have been described here.

We hope that these expressions for the magnetic field will help dispel the notion that exact expressions for the field of an ideal solenoid are necessarily complicated or slow to work with and we hope they will encourage student investigations of magnetic phenomena.
\appendix
\section{Generalized Complete Elliptic Integral}
\label{sec: appcel}
\begin{figure}
    \begin{center}
    \includegraphics[scale=.8]{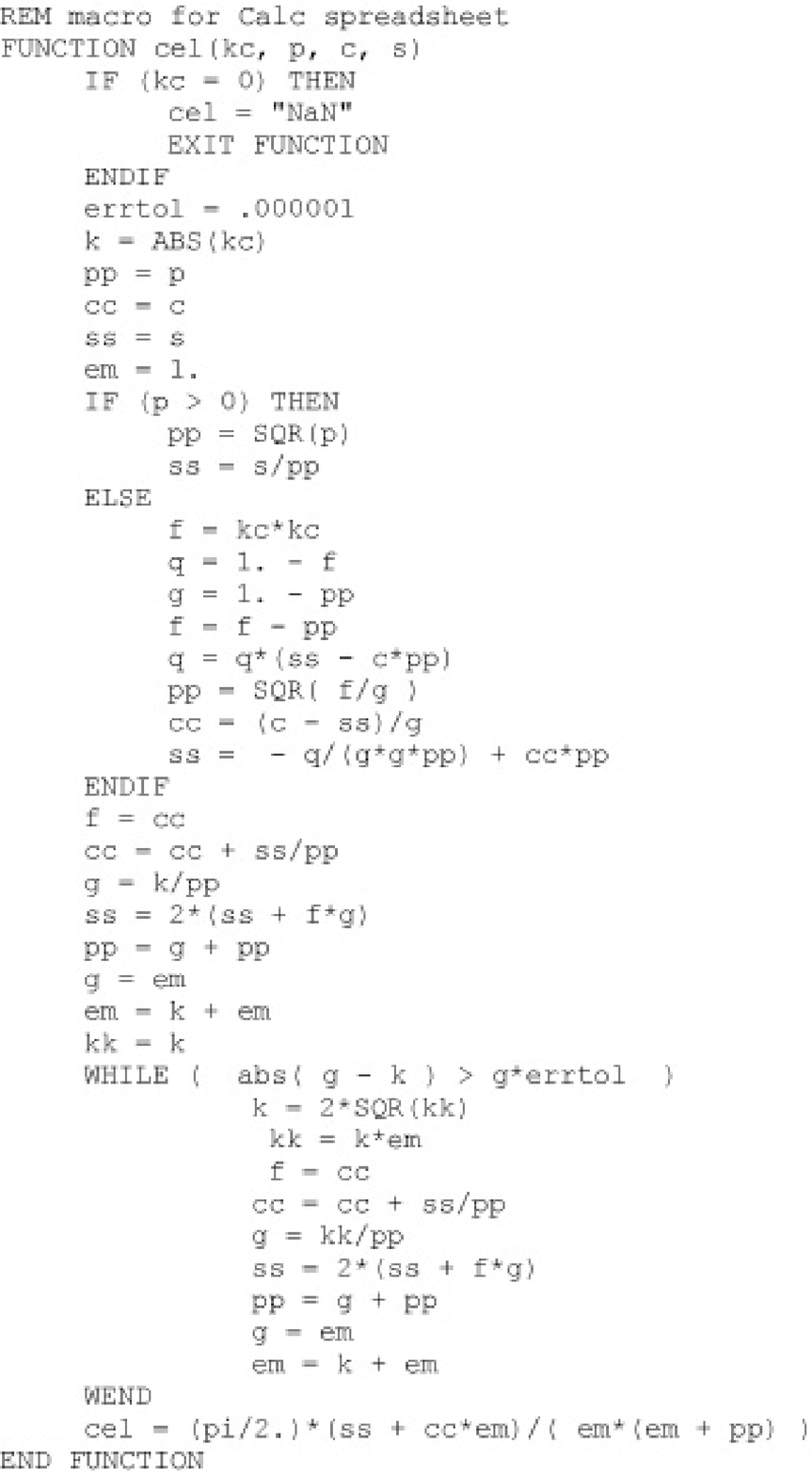}
    \end{center}
    \caption{Algorithm for the generalized complete elliptic integral $C(k_c ,p,c,s)$ coded in a version of BASIC for use as an add-in function in the $Calc$ spreadsheet. } \label{fig:algorithm}
\end{figure}

The generalized complete elliptic integral in Eq.~\eqref{celdef} can be efficiently computed using an algorithm by Bulirsch \cite{Bulirsch} based on work of Bartky \cite{Bartky} who extended ideas of Landen and Gauss. This algorithm converges so quickly that, unless $k_c << 1$, only 3 or 4 passes are necessary. The code shown in Fig. \ref{fig:algorithm} is in a version of BASIC that can be loaded as a user-defined function into the $Calc$ spreadsheet program (part of the free Open Office\cite{openoffice} software suite) where it can then be used as a normal spreadsheet function, making $C$ accessible to non-programmers. Similarly, Microsoft $Excel$ allows user defined function coded in Visual BASIC. This code is simple enough to be treated as pseudocode that can be easily adapted to other languages. Alternatively, code in FORTRAN or C can be found on the internet\cite{fortran} or in the first edition of {\it Numerical Recipes}.\cite{NR} Code in FORTRAN, C, or Python is also available from the authors.
Later editions of {\it Numerical Recipes} no longer mention $cel$ and adopt instead Carlson's more general approach, which applies to incomplete as well as complete elliptic integrals. In terms of Carlson's functions, $R_F$ and $R_C$, $cel$ is:
\begin{equation}
\label{Carlson}
C\left( {k_c ,p,c,s} \right) = c\,R_F \left( {0,k_c ^2 ,1} \right) + \frac{1}{3}\left( {s - pc} \right)\,\,R_J \left( {0,k_c ^2 ,1,p} \right) .
\end{equation}
Code for computing Carlson's functions may be found in the current editions of {\it Numerical Recipes} and elsewhere. However, for {\it complete} elliptic integrals, Bulirsch's algorithm is more compact and also has some other advantages.\cite{compare}
This generalized complete elliptic integral includes all three standard Legendre forms as special cases:
\begin{equation}
\label{Leg1}
K\left( {k} \right) = C\left( {k_c ,1 ,1,1} \right) ,\,\,\,\, E\left( {k} \right) = C\left( {k_c ,1 ,1,k_c ^2} \right) , \,\,\,\,
\Pi\left(  {n,k} \right) = C\left( {k_c ,n+1 ,1,1} \right) ,
\end{equation}
where
\begin{equation}
\label{Leg2}
k = \sqrt {1 - k_c ^2} .
\end{equation}
In algebraic work, the following identity is useful:	
\begin{equation}
\label{recursion}
C\left( {k_c ,\gamma ^2 ,2 - \gamma,\gamma} \right) - C\left( {k_c ,1,1,1} \right) \equiv \left( {1 - \gamma} \right)\,\,C\left( {k_c ,\gamma^2 ,1,\gamma} \right) .
\end{equation}
\section{Derivation of Solenoid Field}
\label{sec: appder}
The magnetic field of an ideal solenoid can be computed directly from the Biot-Savart law. The necessary algebra is only slightly more complicated than that which is needed to derive Eq.~\eqref{finsolax}, which is commonly presented to students in introductory courses. The notation used for $\alpha_\pm, k_\pm, etc.$ is the same as that of section \ref{sec: intro}.
\begin{figure}
    \begin{center}
    \includegraphics[scale=.7]{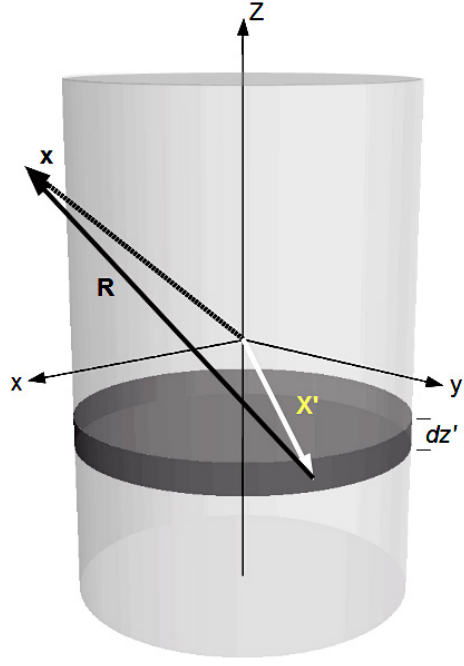}
    \end{center}
    \caption{Geometry of an ideal solenoid showing notation used in applying the Biot-Savart law. }
    \label{fig:maggeo}
\end{figure}

First, the surface of the solenoid is divided into circular strips of width $dz'$ as in Fig.~\ref{fig:maggeo}. The current in such a strip is $nI dz'$. To calculate the field at a point $\bf{x}$, we apply the Biot-Savart law to this circular loop and then add up the fields of the stack of strips that form the solenoid surface:
\begin{equation}
{\bf{B}}\left( {\bf{x}} \right) = \int\limits_{ - b}^b {\,\,\,\,\frac{{\mu _o }}{{4\pi }}\,\,\,\,\oint {\left( {nI\,dz'} \right)\frac{{\,\,d{\bf{x'}} \times {\bf{R}}}}{{\left| {\bf{R}} \right|^3 }}} } ,
\end{equation}
where
\begin{equation}
\bf{R}=\bf{x}-\bf{x'},
\end{equation}
and points along the strip are specified by the position vector
\begin{equation}
{\bf{x'}} = a\cos \varphi '\,{\bf{\hat i}} + a\sin \varphi '\,{\bf{\hat j}} + z'\,{\bf{\hat k}},
\end{equation}
and an infinitesimal step taken along the strip is
\begin{equation}
d{\bf{x'}} = \left( { - a\sin \varphi '\,{\bf{\hat i}} + a\cos \varphi '\,{\bf{\hat j}}} \right)d\varphi ' .
\end{equation}

Because of the cylindrical symmetry, we are free to choose coordinates in which $\bf{x}$ lies in the {\it x-z} plane, causing $B_y(\bf{x})$ to integrate to $0$ and allowing $B_x(\bf{x})$ to be identified with $B_\rho(\bf{x})$, suggesting the notation
\begin{equation}
{\bf{x}} = \rho {\bf{\hat i}} + z\,{\bf{\hat k}} .
\end{equation}

Since
\begin{equation}
{\bf{R}} = \left( {\rho  - a\cos \varphi '} \right){\bf{\hat i}} - a\sin \varphi '{\bf{\hat j}} + \left( {z - z'} \right)\,{\bf{\hat k}} ,
\end{equation}
and
\begin{equation}
d{\bf{x'}} \times {\bf{R}} = a\,d\varphi '\left[ {\left( {z - z'} \right)\cos \varphi '{\bf{\hat i}} + \left( {z - z'} \right)\sin \varphi '{\bf{\hat j}} + \,\left( {a - \rho \cos \varphi '} \right){\bf{\hat k}}} \right] ,
\end{equation}
the field can be written as
\begin{equation}
{\bf{B}}\left( {\bf{x}} \right) = \int\limits_{ - b}^b {\,\,dz'\,\,\left( {\frac{{\mu _o nIa}}{{2\pi }}} \right)\,\,\,\int\limits_0^\pi  {d\varphi '\frac{{\left( {z - z'} \right)\cos \varphi '{\bf{\hat i}} + \left( {a - \rho \cos \varphi '} \right){\bf{\hat k}}}}{{\left\{ {\rho ^2  - 2a\rho \cos \varphi ' + a^2  + \left( {z - z'} \right)^2 } \right\}^{{\raise0.7ex\hbox{$3$} \!\mathord{\left/
 {\vphantom {3 2}}\right.\kern-\nulldelimiterspace}
\!\lower0.7ex\hbox{$2$}}} }}} \,} .
\end{equation}

Integration over $z'$ is elementary:
\begin{equation}
\label{BrhiInt}
B_\rho  \left( {\bf{x}} \right) = \,\,\left( {\frac{{- B_o a}}{2}} \right)\int\limits_0^\pi  {d\varphi '\cos \varphi '\left\{ {\frac{1}{{\sqrt {z_ +  ^2  + \rho ^2  + a^2  - 2a\rho \cos \varphi '} }} - \frac{1}{{\sqrt {z_ -  ^2  + \rho ^2  + a^2  - 2a\rho \cos \varphi '} }}} \right\}}
\end{equation}
and
\begin{equation}
\label{BzInt}
B_z \left( {\bf{x}} \right) = \,\,\left( {\frac{{B_o a}}{2}} \right)\int\limits_0^\pi  {d\varphi '\frac{{\left( {a - \rho \cos \varphi '} \right)}}{{\left( {\rho ^2  + a^2  - 2a\rho \cos \varphi '} \right)}}\left\{ \begin{array}{l}
 \frac{{z_ +  }}{{\sqrt {z_ +  ^2  + \rho ^2  + a^2  - 2a\rho \cos \varphi '} }} \\
  - \frac{{z_ -  }}{{\sqrt {z_ -  ^2  + \rho ^2  + a^2  - 2a\rho \cos \varphi '} }} \\
 \end{array} \right\}} .
\end{equation}
	
To put these expressions into a form resembling $cel$, introduce a change in integration variable:
\begin{equation}
\label{psi}
2\psi  \equiv \pi  - \varphi '
\end{equation}
and, after using some trigonometric identities, observe that
\begin{equation}
\label{kpmdenom}
z_ \pm  ^2  + \rho ^2  + a^2  - 2a\rho \cos \varphi ' = \left[ {z_ \pm  ^2  + \left( {\rho  + a} \right)^2 } \right]\left( {\cos ^2 \psi  + k_ \pm  ^2 \sin ^2 \psi } \right) .
\end{equation}

The radial component of $\bf{B}$ then becomes
\begin{equation}
B_\rho  \left( {\bf{x}} \right) = \,\,B_o \int\limits_0^{{\raise0.7ex\hbox{$\pi $} \!\mathord{\left/
 {\vphantom {\pi  2}}\right.\kern-\nulldelimiterspace}
\!\lower0.7ex\hbox{$2$}}} {d\psi \left( {\cos ^2 \psi  - \sin ^2 \psi } \right)\left\{ {\frac{{\alpha _ +  }}{{\sqrt {\cos ^2 \psi  + k_ +  ^2 \sin ^2 \psi } }} - \frac{{\alpha _ -  }}{{\sqrt {\cos ^2 \psi  + k_ -  ^2 \sin ^2 \psi } }}} \right\}} .
\end{equation}

Upon comparing each of the two terms in this integrand to the integrand in the definition of $cel$, we recognize that the radial field can be identified as Eq.~\eqref{Brho}.

Similarly, the longitudinal component of the field becomes
\begin{equation}
B_z \left( {\bf{x}} \right) = \,\,\frac{{B_o a}}{{\left( {\rho  + a} \right)}}\int\limits_0^{{\raise0.7ex\hbox{$\pi $} \!\mathord{\left/
 {\vphantom {\pi  2}}\right.\kern-\nulldelimiterspace}
\!\lower0.7ex\hbox{$2$}}} {d\psi \frac{{\cos ^2 \psi  + \gamma \sin ^2 \psi }}{{\cos ^2 \psi  + \gamma ^2 \sin ^2 \psi }}\left\{ {\frac{{\beta _ +  }}{{\sqrt {\cos ^2 \psi  + k_ +  ^2 \sin ^2 \psi } }} - \frac{{\beta _ -  }}{{\sqrt {\cos ^2 \psi  + k_ -  ^2 \sin ^2 \psi } }}} \right\}} ,
\end{equation}
which can be recognized as Eq.~\eqref{Bz}.

In order to determine the magnetic field of a current loop ($0 < b << a$), it is simplest to return to Eq.~\eqref{BrhiInt} and Eq.~\eqref{BzInt}, and, treating $b$ as a small quantity, expand the integrands to first order in $b$ and then repeat the transformations in Eq.~\eqref{psi} and Eq.~\eqref{kpmdenom} to put the integrals in forms resembling $cel$. The results are given in Eq.~\eqref{Brholoop} and Eq.~\eqref{Bzloop}.

\begin{acknowledgments}
The authors wish to acknowledge the support and assistance of John Belcher  and the Center for Educational Computing Initiatives at the Massachusetts Institute of Technology.
\end{acknowledgments}

$^{a)}$ Electronic address:  nderby1@comcast.net

$^{b)}$ Electronic address:  stanolbert@comcast.net

\end{document}